# Evaluating the Effects of Control Surfaces Failure on the GTM

Ramin Norouzi[a], Amirreza Kosari[a,*], and Mohammad Hossein Sabour[a]

[a] *Aerospace Department, Faculty of New Sciences and Technologies, University of Tehran, Tehran, Iran*

**Abstract**

Despite the advances in aircraft guidance and control systems technology, Loss of Control remains as the main cause of the fatal accidents of large transport aircraft. Loss of Control is defined as excursion beyond the allowable flight envelope and is often a consequence of upset condition if improper maneuver is implemented by the pilot. Hence, extensive research in recent years has focused on improving the current fault tolerant control systems and developing new strategies for loss of control prevention and recovery systems. However, success of such systems requires the perception of the damaged aircraft's dynamic behavior and performance, and understanding of its new flight envelope. This paper provides a comprehensive understanding of lateral control surfaces' failure effect on the NASA Generic Transport Model's maneuvering flight envelope; which is a set of attainable steady state maneuvers herein referred to as trim points. The study utilizes a massive database of the Generic Transport Model's high-fidelity maneuvering flight envelopes computed for the unimpaired case and wide ranges of aileron and rudder failure cases at different flight conditions. Flight envelope boundary is rigorously investigated and the key parameters confining the trim points at different boundary sections are identified. Trend analysis of the impaired flight envelopes and the corresponding limiting factors is performed which demonstrates the effect of various failure degrees on the remaining feasible trim points. Results of the post-failure analysis can be employed in emergency path planning and have potential uses in the development of aircraft resilient control and upset recovery systems.

## 1. Introduction

Based on the statistical report published by Boeing in October 2018, Loss of Control (LOC); with 14 accidents and a total of 1129 fatalities, is the primary contributor among different factors leading to fatal accident of commercial airliners over the years 2008-2017 [1]. Also, the report published by UK Civil Aviation Authority in 2013 investigating the fatal accidents of 2002-2013 indicates that almost 40% of all fatal accidents were related to the loss of control, making it the major cause of the accidents [2].

Despite the increase in the number of flights, there has been a decreasing trend in the number of fatal accidents which is mainly due to the emerging of more accurate and intelligent flight management and control automation systems [2]. However, LOC still holds the greatest share in the fatal accidents.

LOC usually occurs following an upset condition which is caused by external events such as icing, or technical failures such as control surface defects, or internal sources such as pilot inputs, or a combination of these factors [3].

In the case of technical failures, aircraft dynamics and parameters are changed, and so are the flight envelope and its kinematic constraints. As the pilots are not and cannot be trained for all possible failure cases, the aircraft altered dynamics and the new flight envelope boundary are not determined for the pilot. So, the pilot who tries to plan a safe landing trajectory may apply a control input which yields a maneuver outside the new admissible flight envelope leading to LOC [4].

Hence, in order to prevent LOC-led-accidents, it is crucial to increase the pilot's situational awareness and develop better control systems that comply with the dynamics alteration, which both require comprehending the effects of the damage on the aircraft's flight envelope [5]. To address this requirement, flight envelopes of the unimpaired and the damaged aircraft must be computed and thoroughly investigated to identify the parameters which shape the flight envelope boundary and to evaluate their variations with different degrees of failure.

Various approaches have been used in the previous studies to estimate the flight envelope. As mentioned earlier, maneuvering flight envelope (MFE) is a set of attainable trim states within a set of constraints and loss of control may occur once any of the constraints is violated [6]. Being attainable means the trim state belongs to the region of attraction of the equilibrium state and the new stability margin is adequate for the impaired aircraft to cope with disturbances such as gust

---

* Correspondence should be addressed to Amirreza Kosari; kosari_a@ut.ac.ir

[7]. Hence, one method is to determine the region of attraction of the equilibrium points in the nonlinear system. For instance, in [8], Lyapunov function method has been used to estimate the attraction region of a stable equilibrium point in a nonlinear system. Also in [9], Linear reachable set and nonlinear region of attraction techniques were used to develop algorithms to assess the dynamic flight envelope of the NASA Generic Transport Model (GTM). More recently, the region of attraction representing the dynamic flight envelope was constructed using the stable manifold theory [10, 11].

Another approach is to calculate the reachability-based safe envelope by transforming the problem into Hamilton-Jacobi partial differential equations and solving it with level set methods [12, 13, 14, and 15].

There are also other researches in which the flight envelope is estimated by directly evaluating the trim states in different conditions. For instance, in [16], turning, pull-up, and push-over steady state conditions are derived by numerically minimizing a cost function, whereas in [17], interval analysis method is used to derive trim states. In [18], steady states are computed using the Newton-Raphson method to evaluate the unimpaired aircraft's steady performance and maneuvering capabilities in helical trajectories. In [19], similar problem has been addressed in which, instead of trimming every steady state point, boundary of the impaired aircraft's flight envelope is directly computed using the continuation technique. Specifically, the approach of computing 3D maneuvering flight envelopes of an impaired aircraft by evaluating all trim points in a point-by-point schema was introduced in [20] and elaborated in [21]. The method characterizes the trim points by velocity, climb rate, turn rate and altitude, and derives them by simultaneously minimizing the 6 degree-of-freedom (DOF) nonlinear equations of motion according to the altered dynamics of the damaged aircraft. This approach has been used in several studies associated with flight envelope estimation of impaired aircraft and post-failure path planning such as [7, 22, and 23]. Also in [24, 25, and 26], this method was applied to the NASA GTM with left wing damage to estimate the post-failure maneuvering flight envelopes.

Also, researchers concerned with estimating the impaired aircraft's flight envelope in real-time have adopted various approaches. For instance, in [27, 28-30], the intended flight envelope is estimated by interpolating closely related envelopes retrieved from an offline generated database being carried onboard. Also, a neural network-based method has been proposed in [5] which estimates the boundary of the impaired aircraft's global flight envelope in real-time.

Despite many studies devoted to the flight envelope estimation and those investigating the envelope characteristics based on limited failure cases and in local ranges of the flight envelope, to the best of our knowledge; so far there has been no comprehensive study on the flight envelope variations due to wide ranges of failure degrees. In this research, boundaries of the unimpaired and impaired maneuvering flight envelopes of the NASA GTM are rigorously analyzed to investigate the envelope variations with different degrees of the control surfaces failure and various flight conditions.

The rest of this paper is organized as follows: in Sec. 2, the GTM which is the dynamic model used in this research is introduced. In Sec 3, trim state definition, stability evaluation, and the computational procedure by which the utilized MFE database was generated are explained, and Sec. 4 presents the specifications of the database. The analysis performed on the maneuvering flight envelopes is presented in Sec. 5, where the results of the rigorous investigation of the unimpaired flight envelope and its limiting factors are discussed in subsection 5.1, and the variations of the flight envelope with different degrees of the control surfaces failure at a number of flight conditions are presented and discussed in subsection 5.2. Finally, Sec. 6 concludes the paper.

## 2. NASA GENERIC TRANSPORT MODEL

The Generic Transport Model (GTM) – with tail number T2 is a 5.5%, dynamically scaled, twin engine aircraft which resembles a commercial jet airliner and is designed to be flown into drastic upset conditions and being safely recovered [31]. The GTM-T2 specifications are shown in Table 1.

TABLE 1
GTM-T2 PROPERTIES

| Property | Quantity |
|---|---|
| Takeoff weight, $W_0$ | 257 N (26.2 kg) |
| Wing area, S | 0.5483 m$^2$ |
| Wing span, b | 2.09 m |
| Length, l | 2.59 m |
| Mean aerodynamic chord, $\bar{c}$ | 0.2790 m |

The MFE database employed in this research was created using the high fidelity, nonlinear, 6 DOF, MATLAB® – Simulink® model of the GTM-T2 known as *"GTM-DesignSim"* [32]. The model utilizes an extended-envelope aerodynamic dataset which was created by performing extensive wind tunnel tests on the GTM at angles of attack and sideslip angles ranging from −5° to +85° and −45° to +45°, respectively [31]. Further details on the GTM can be found in [31, 33-37].

## 3. FLIGHT ENVELOPE ESTIMATION

In this section, the process in which maneuvering flight envelopes of the database were estimated is briefly described. More detailed description can be found in [4, 5].

### 3.1. Trim state

As mentioned earlier, flight envelopes of the previously generated database which is used in this research are actually maneuvering flight envelopes, which mean they are boundaries containing steady state maneuvers. Such steady state maneuvers are referred to as trim points in this research.

A steady state maneuver is the condition in which all linear and angular velocity rates and aerodynamic angles rates are zero [38]. Hence, in the wind-axes coordinate system:

$$(\dot{V}, \dot{\alpha}, \dot{\beta}) = (\dot{p}, \dot{q}, \dot{r}) = 0 \qquad (1)$$

where, $V$, $\alpha$ and $\beta$ are the aircraft total airspeed, angles of attack and sideslip, and $p, q, r$, are the roll rate, pitch rate and yaw rate, respectively.

Since the considered steady state maneuvers are in fact level-climbing-descending rectilinear and turning flights,

$$\dot{\phi}, \dot{\theta} = 0, \quad \gamma = \gamma^*, \quad \dot{\psi} = \dot{\psi}^* \tag{2}$$

where, $\phi, \theta, \psi$, and $\gamma$ are the roll (bank) angle, pitch angle, yaw (heading) angle, and the flight path angle. Also, $\gamma^*, \dot{\psi}^*$ are the desired constant values defining the steady state maneuvers.

In terms of the nonlinear aircraft equations of motion:

$$\dot{x}_{trim} = f(x_{trim}, u_{trim}) = 0 \tag{6}$$

$$x_{trim} = [V, \alpha, \beta, p, q, r, \phi, \theta]^T = x^* \tag{7}$$

$$u_{trim} = [\delta_{th}, \delta_e, \delta_a, \delta_r]^T = u^* \tag{8}$$

where $f$ is a vector of nonlinear functions, $x$ is the state vector, $u$ is the control vector, and $\delta_{th}$, $\delta_e$, $\delta_a$, and $\delta_r$ represent engine throttle setting and deflections in the elevator, aileron, and rudder respectively.

Maneuvering flight envelopes are comprised of trim states characterized by four parameters $(h^*, V^*, \gamma^*, \dot{\psi}^*)$. These flight envelopes can be depicted as three-dimensional volumes at each constant flight altitude $h^*$ [4]. Hence, each trim point should be derived by solving all the aircraft nonlinear equations of motion ($\dot{x}_{trim} = 0$) for the intended flight path angle and turn rate $(\gamma^*, \dot{\psi}^*)$ at the considered airspeed $(V^*)$ and altitude $(h^*)$.

*3.2. Computational procedure*

Since this is not analytically possible, the corresponding constrained nonlinear optimization problem is numerically solved, in which the cost function is [38, 21, and 24]:

$$J(x, u) = \frac{1}{2} \dot{x}_{trim}^T Q \dot{x}_{trim} \tag{17}$$

where $Q$ is a positive definite weighting matrix which specifies the contributions of the state derivatives to the cost function. $J$ is subject to the following equality and inequality constraints:

$$h - h^* = 0 \tag{18}$$

$$V - V^* = 0 \tag{19}$$

$$tan\theta - \frac{ab + sin\gamma^* \sqrt{a^2 - sin^2\gamma^* + b^2}}{a^2 - sin^2\gamma^*} = 0, \quad \theta \neq \pm\pi/2 \tag{20}$$

where, $a = cos\alpha cos\beta, \quad b = sin\phi sin\beta + cos\phi sin\alpha cos\beta \tag{21}$

$$p + \dot{\psi}^* sin\theta = 0 \tag{22}$$

$$q - \dot{\psi}^* cos\theta sin\phi = 0 \tag{23}$$

$$r - \dot{\psi}^* cos\theta cos\phi = 0 \tag{24}$$

$$\begin{aligned} |\delta_{th} - 0.5| \leq 0.5 & \quad |\delta_e| \leq 30 \\ |\delta_a| \leq 20 & \quad |\delta_r| \leq 30 \end{aligned} \tag{26}$$

To construct the database used in this research, the aforementioned constrained optimization problem was iteratively solved via the sequential quadratic programming (SQP) technique with a convergence criterion of $10^{-7}$.

*3.3. Linearization and stability evaluation*

At the final step, the stability of the derived feasible trim points are checked as being feasible is not adequate for a trim state to include it in the maneuvering flight envelope and being stable is the sufficient condition.

By definition, a nonlinear system is considered stable at a specific trim point if it inherently converges to that trim state when being in the vicinity of the trim point [39].

To evaluate the stability of the aircraft at a derived trim point $x^*$, the aircraft equations of motion are linearized about $x^*$ via a perturbation method:

$$\dot{X} = AX + BU \tag{11}$$

where, $X = x - x^*$, $U = u - u^*$, and $A, B$ are constant Jacobian matrices:

$$A = \left. \frac{\partial f}{\partial x} \right]_{x^*, u^*} \tag{12}$$

$$B = \left. \frac{\partial f}{\partial u} \right]_{x^*, u^*} \tag{13}$$

System eigenvalues determine if the considered trim point is stable. While stability is more preferable as the aircraft naturally tends to damp the effect of small disturbances, an unstable trim point still can be included in the flight envelope if it is controllable, which is evaluated by the linearized system's controllability matrix $C$:

$$C = [B \quad AB \quad A^2B \quad \cdots \quad A^{n-1}B] \tag{16}$$

*3.4. Additional considerations*

*3.4.1. Bank angle*

According to section 1, one of the main reasons for the impaired aircraft's maneuvering flight envelope estimation is the post-failure path planning. The load factor $(n)$ becomes larger with increase in airplane's bank angle, and the turn becomes less comfortable and more annoying for passengers:

$$n = 1/cos\phi \tag{27}$$

Hence, it is more preferable for the commercial transport aircraft to use bank angles of less than 30 degrees in the path planning, especially in the final approach and landing. This enables shallow turns which impose small $g$-forces (up to 1.2$g$) on passengers. For instance, in [24], an adaptive flight planner which selects trim points from the estimated flight envelopes was applied to a wing damaged GTM for the post-failure path planning in the landing phase and a 30° bank constraint was imposed on the path planner to have shallow turns to the final approach. Also, in [25], maneuvering flight envelopes were estimated for bank angles constrained to ±20°. Likewise in [40], a 35° bank constraint was imposed on the online flight envelope determination of an Airbus A300 full-scale model. Therefore, a 30° bank constraint was imposed on the trim points derivation of the utilized database. This way, the estimated MFEs are composed of trim points which all require bank angles equal to or less than 30°. It should be noted that such constraint does not mean that the aircraft is incapable of flying with banks more than 30°. However, it was imposed to derive trim points that are all possible to be implemented by the flight management system and the autopilot, and also to investigate how far the aircraft can maneuver satisfying this constraint. Perhaps, other researchers can estimate the GTM flight envelopes with bank angles up to 67° structural limit or up to any other degree based on their research goal.

*3.4.2. Angle of attack*

As mentioned in section 2, the GTM was designed to investigate the stall and post-stall regimes in the extended flight envelope regions, so the model's aerodynamic dataset includes aerodynamic data for high values of the angle of attack up to +85°. However, in this research, we intend to estimate normal operating flight envelopes for the unimpaired and impaired cases. Therefore, we need to impose an angle of attack constraint to prevent the optimization algorithm from trimming the aircraft at the stall or post-stall flight conditions. To do so, we need to consider the critical angle of attack, at which, the airplane has maximum lift coefficient. This angle corresponds to the stall speed. For velocities below this speed or angles of attack above this angle, the flow separates from wings and the aircraft loses the required lift. According to the results of the bifurcation analysis of the GTM presented in [3], from $\alpha = 10.5°$, the aircraft enters an undesirable regime with steep helical spirals which are considered as upset conditions and must be recovered from, by stall recovery procedures. Also, in [41], the angle of attack is restricted to 10° such that no part of the aircraft stalls. Therefore, the employed database was constructed with the angle of attack constrained to 10.5°.

## 4. MANEUVERING FLIGHT ENVELOPE DATABASE

As mentioned earlier, a previously generated database [42] comprising the flight envelopes of the unimpaired and impaired GTM is used to perform this study. The 3D MFEs of the database were evaluated by solving the presented computational procedure in an iterative scheme in which for every new trim point, the initial guess for the solution of the optimization is set equal to the successful solution of the optimization for the previous trim point. This way, the convergence of the algorithm for every feasible trim point is almost guaranteed, and the probability of being trapped in local minima becomes very low.

In this study, we aim to investigate the trend of changes in the maneuvering flight envelopes and their boundaries due to various failure degrees of aileron and rudder. There are four categories of control surface failures [43]:

- Control restriction, in which, upper and/or lower limits of deflection are changed to new, equal or non-equal values
- Surface jam
- Reduced rate limits, in which, upper and/or lower rate limits are changed
- Surface runaway, which at first shows up as reduced rate limits, but eventually changes to the surface jam case

Control restriction and surface jam are caused by physical damage, icing, or a loss of hydraulic power. The fourth category eventually becomes surface jam; hence there are three main failure categories which two of them have been considered in the database: control restriction and surface jam.

In order to have a comprehensive database, different sections of the aileron and rudder operational ranges were covered by selecting the failure degrees from lowest to highest as presented in Table 2. It should be noted that the values in the brackets are the lower limit ($LL$) and the upper limit ($UL$) of the control surface deflection, as in [$LL, UL$]. Thus, jamming failure is a special case of the restriction failure, in which $LL$ and $UL$ are identical. Also, in the investigation process of the failure trim points, the limits in the inequality constraints (25) were set to the corresponding $LL$ and $UL$ of the failures.

TABLE 2
GTM-T2 CONTROL SURFACE FAILURES

| Failure Type | Control Surface | Failure |
|---|---|---|
| Surface Jam | Aileron | $-20°, -10°, 0°, 10°, 20°$ |
| | Rudder | $-30°, -20°, -10°, 0°, 10°, 20°, 30°$ |
| Control Restriction | Aileron | $[-20°, -10°], [-20°, 0°], [-20°, 10°], [-10°, 0°], [-10°, 10°], [10°, 20°], [0°, 20°], [-10°, 20°], [0°, 10°]$ |
| | Rudder | $[-30°, -20°], [-30°, -10°], [-30°, 0°], [-30°, 10°], [-30°, 20°], [20°, 30°], [10°, 30°], [0°, 30°],$ $[-10°, 30°], [-20°, 30°], [-20°, 20°], [-20°, -10°], [-20°, 0°], [-20°, 10°], [-10°, 0°], [-10°, 10°]$ $[10°, 20°], [0°, 20°], [-10°, 20°], [0°, 10°]$ |

TABLE 3
3D MANEUVERING FLIGHT ENVELOPE

| Failure Type | Control Surface/Altitude | Quantity | | |
|---|---|---|---|---|
| Surface Jam | Aileron | 5 | | |
| | Rudder | 7 | 48 | |
| | Altitude | 4 | | 168 |
| Control Restriction | Aileron | 9 | | |
| | Rudder | 20 | 116 | |
| | Altitude | 4 | | |
| Unimpaired | Altitude | 4 | 4 | |

To enable assessing the effect of different flying altitudes, the 3D MFEs of the database have been estimated at four different altitudes of Sea Level, 10000 ft, 20000 ft, and 30000 ft. Hence, the database contains 3D MFEs of 168 different cases (Table 3).

Since GTM is symmetrical about the longitudinal axis, MFE of the impaired case [$LL, UL$] is symmetric to the MFE of the impaired case [$\min(-LL, -UL), \max(-LL, -UL)$]. For instance, MFE of the impaired case with rudder restricted to [−30°, 10°], is symmetry of the MFE of the impaired case with rudder restricted to [−10°, 30°]. Therefore in the evaluation of the 3D MFEs, flight envelopes of almost half of the failure cases were evaluated based on their symmetricals. For such cases, flight envelope boundaries were validated by checking randomly selected trim points from inside and outside of the boundary. It should be noted that the control surface sign conventions used in the GTM model are:

TABLE 4
CONTROL SURFACE SIGN CONVENTIONS

| Control Surface | Positive Deflection |
|---|---|
| Elevator | Trailing edge down |
| Rudder | Trailing edge left |
| Aileron | Left wing trailing edge up |

The smaller the resolution increments in $V, \gamma,$ and $\dot{\psi}$ ranges, the more the trim points, and hence the higher the accuracy of the flight envelopes and their boundaries. Therefore these increments were selected as per Table 5 so that high fidelity flight envelopes could be estimated. MFEs of the database were evaluated for different flight path angles within the range of $-5° \leq \gamma \leq 5°$. Thus, considering the 1 $deg$ resolution increment in the $\gamma$ range, each 3D MFE is composed of 11 different $\gamma$-constant 2D MFEs $(V - \dot{\psi})$.

TABLE 5
FLIGHT ENVELOPE INCREMENTS

| Parameter | Resolution Increment Size |
|---|---|
| $V$ | 1 knot |
| $\gamma$ | 1 deg |
| $\dot{\psi}$ | 0.2 deg/s |

To construct the database, all computations were split over two standard desktop PCs with 3.00 GHz AMD Phenom quad-core processor, under Windows 7 operating system, and were performed using MATLAB® and Simulink® version 8.2 (R2013b). On average; it took 8 seconds for each trim point to complete the presented numerical process. Therefore, the database was generated by investigating more than 8.8 million trim points in more than 19600 hours in over more than 16 months.

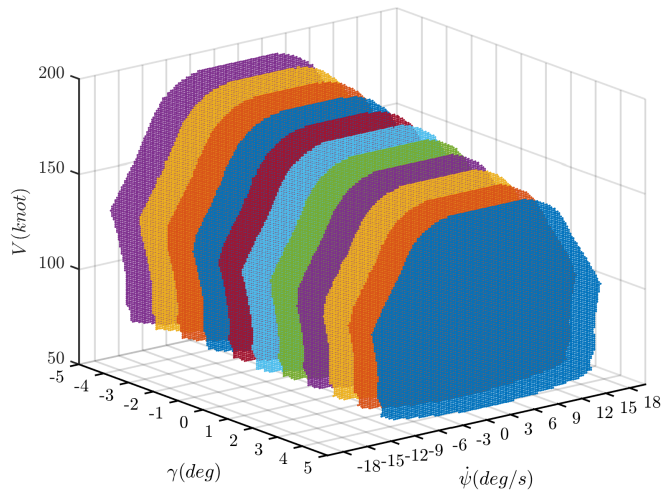

Fig. 1. Unimpaired case at sea level

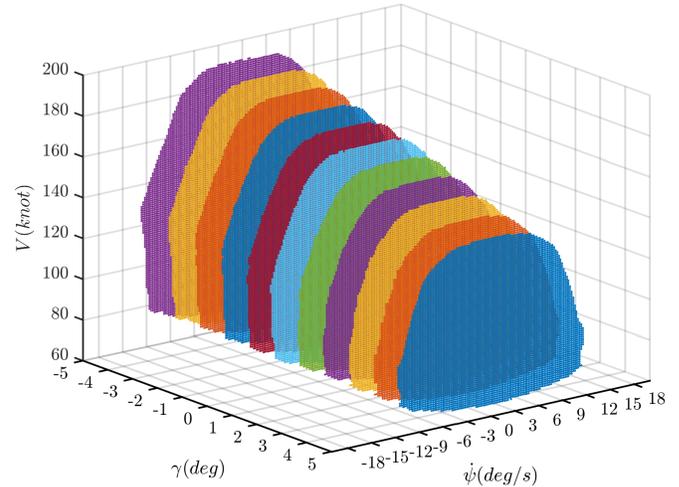

Fig. 3. Unimpaired case at 10000 ft

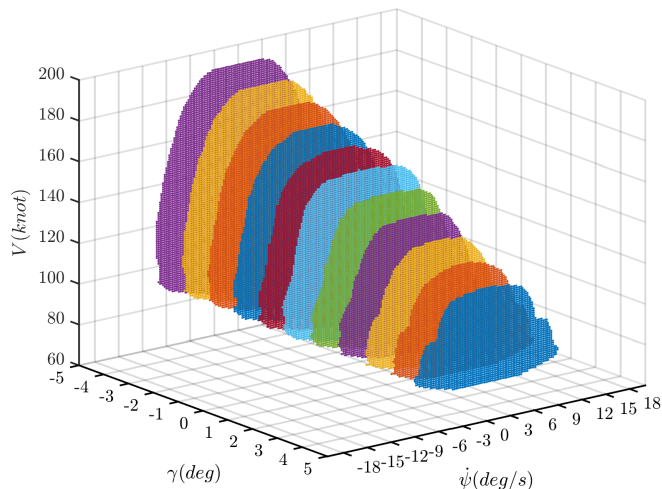

Fig. 2. Unimpaired case at 20000 ft

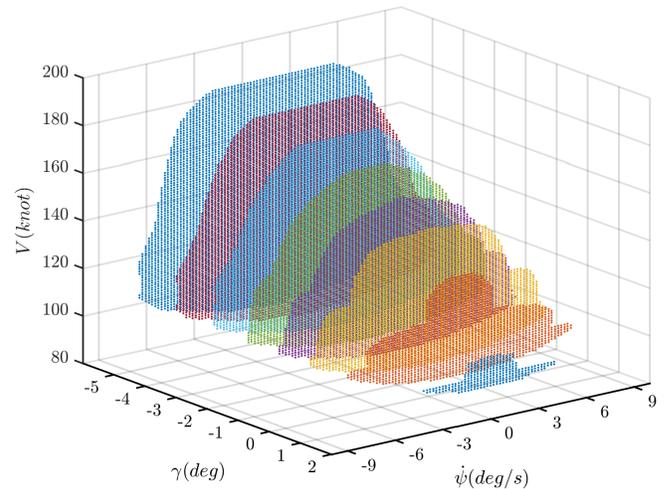

Fig. 4. Unimpaired case at 30000 ft

## 5. ANALYSIS RESULTS AND DISCUSSIONS

### 5.1. Unimpaired flight envelopes

Figures 1 through 4 show 3D maneuvering flight envelopes estimated for the unimpaired GTM in various flight path angles and flying altitudes. These are 4 out of the 168 3D MFEs estimated for the GTM. The rest 164 3D MFEs, which belong to the considered impaired cases, are presented in [42].

Fig. 5 shows the 2D maneuvering flight envelope of the unimpaired GTM at sea level and zero flight path angle. Each blue dot represents a feasible – stable or feasible – controllable trim point. It also shows different sections of the unimpaired flight envelope boundary highlighted by their limiting factors. By limiting factor we mean a determinative parameter which hampers the access to further trim points. Such limiting factors are either a state or a control. Identifying these limiting factors and understanding their variation trends; helps in better comprehension and prediction of changes in flight envelopes due to failures, and better resilient controller design. Hence, in this section, we present the results of a rigorous analysis made on the unimpaired flight envelope boundary of the GTM.

In Fig. 5, the magenta colored section is the stall boundary, in which, no further trim point is admissible, because further trim points would have angles of attack higher than the 10.5° critical angle of attack. Therefore in this section, the limiting factor is the angle of attack. The green colored section represents the aileron saturation boundary, where further trim points would require aileron deflections greater than +20° or −20°. This section's limiting factor is the aileron's deflection angle. The red colored section shows thrust saturation boundary, which is comprised of trim points with maximum throttle setting. Further trim points would require amounts of thrust bigger than the engines' available thrust.

To accurately understand what is going on in these boundaries and why they have these limiting factors, we investigated the states and controls variations along the trim points of these boundaries.

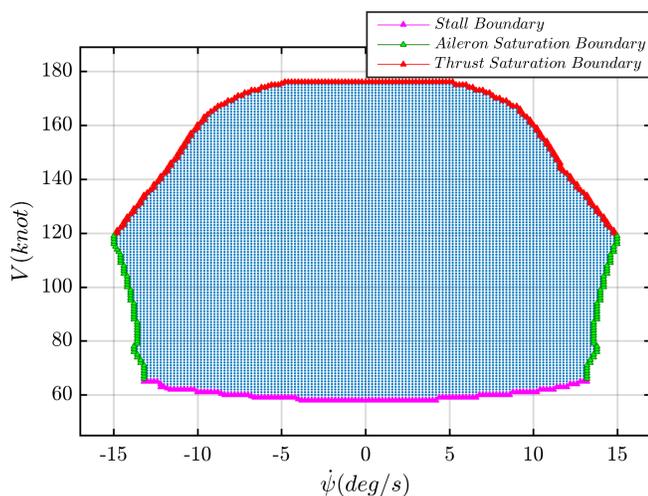

Fig. 5. Flight envelope boundary sections for unimpaired case at sea level and zero flight path angle

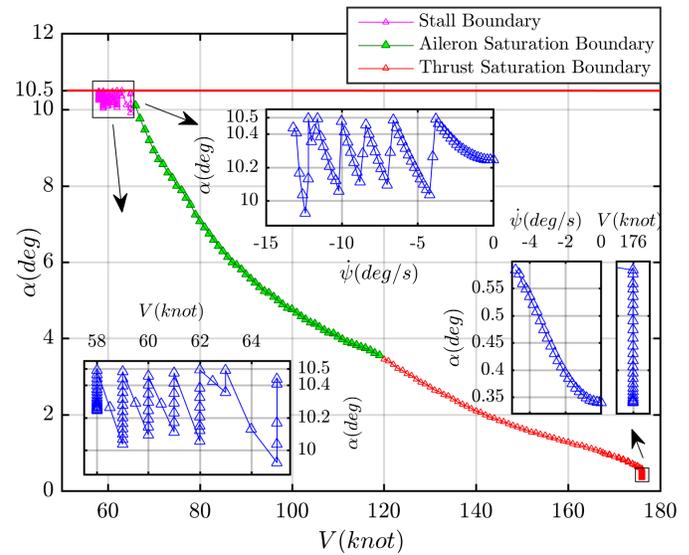

Fig. 6. Variation of angle of attack with boundary trim points

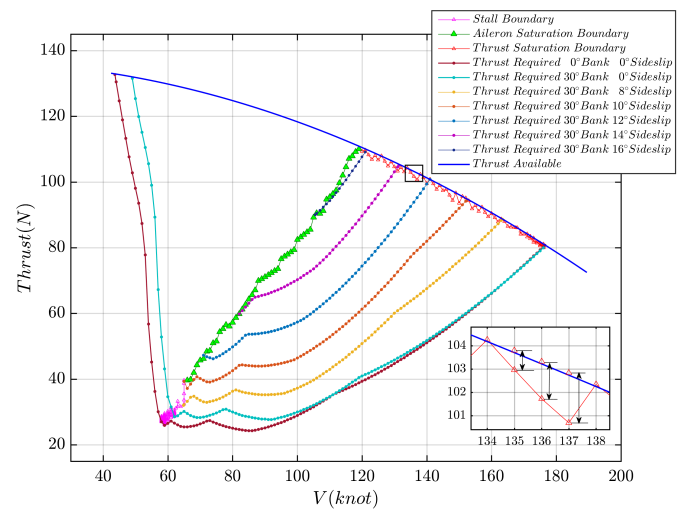

Fig. 7. Variation of thrust-required with boundary trim points

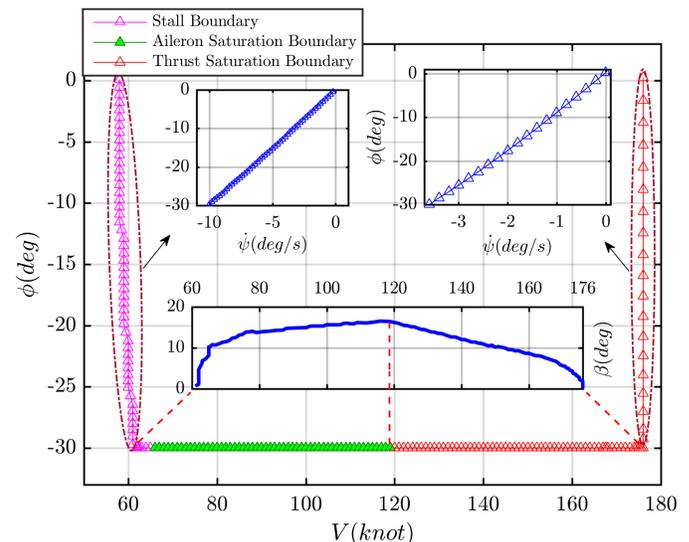

Fig. 8. Variation of bank angle with boundary trim points

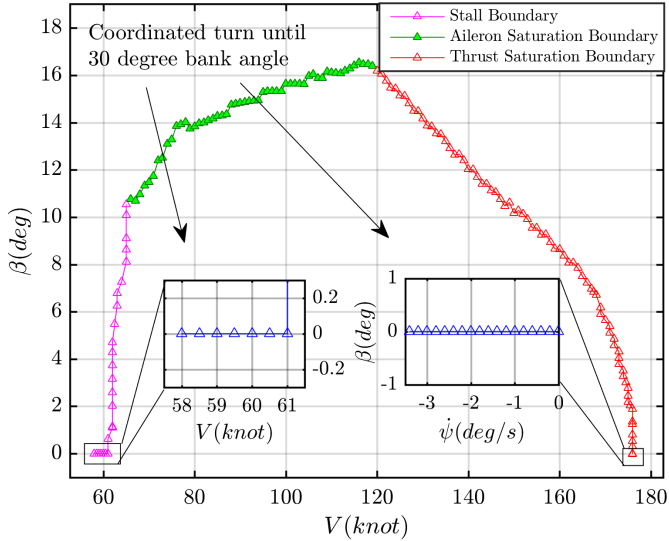

Fig. 9. Variation of angle of sideslip with boundary trim points

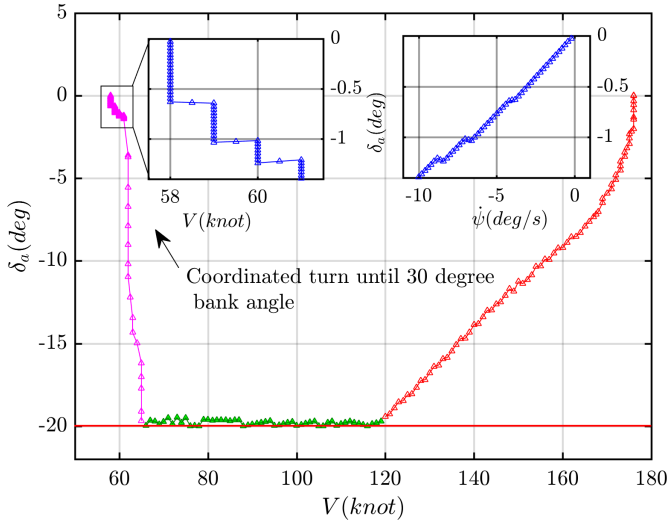

Fig. 10. Variation of aileron deflection angle with boundary trim points

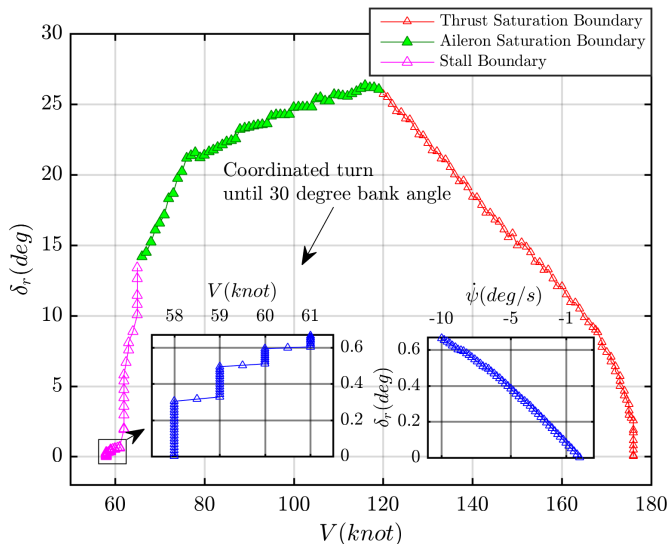

Fig. 11. Variation of rudder deflection angle with boundary trim points

To do so, we started from the rectilinear non-sideslipping maneuver of the lowest speed in the flight envelope, and we moved on the left half of the flight envelope boundary, up to the rectilinear non-sideslipping maneuver of the highest speed in the flight envelope (i.e. from $V = 58, \dot{\psi} = 0$ to $V = 176, \dot{\psi} = 0$). Due to symmetry of the flight envelope, results of this investigation are valid for the other half of the boundary. Figures 6 to 11 show variations of $\alpha$, $T$ (which is the required thrust), $\phi$, $\beta$, $\delta_a$, $\delta_r$, along the trim points of the flight envelope boundary. In each figure, the variation of a state or a control with the flight envelope boundary speed is plotted. Since speed is increasing almost linearly along trim points in most parts of the boundary, speed is chosen as the horizontal axis of these plots, rather than turn rate. An exception to this is the stall boundary section, in which, the primary changing variable is turn rate, whereas speed is constant with respect to different values of turn rate in many sections of the stall boundary. This exception has resulted in congestions in the stall boundary section of most of the presented plots. To enhance the visualization of these plots, zoomed-in-plots of the congested areas are provided inside each figure. Since these zoomed-in-plots are mainly composed of $V$-constant sections, to have a better insight, the corresponding variations in turn rate are also plotted next to the in-plots (in blue), for the same trim points.

By moving from the starting point of the boundary analysis (at the bottom of the envelope) to the end point (at the top of the envelope), and investigating figures 6 to 11 attentively, we can describe the airplane's behavior as below:

The airplane initiates coordinated (zero sideslip) left turn (negative turn rates) via banking to the left (negative roll angle). To increase turn rates, it banks more to the left until the bank angle is saturated (30° bank constraint). From this point onwards, to access higher turn rates, the airplane enters slipping (non-coordinated) turn in which it yaws to the left (increases sideslip) via increasing positive rudder input. However, this would result in an adverse negative roll which increases bank angle if not compensated. Therefore, the airplane counteracts the adverse roll and maintains 30° degree bank angle through increasing negative aileron input. The airplane achieves higher turn rates in this manner until the aileron becomes saturated (end of the stall boundary, beginning of the aileron saturation boundary). At this point, the airplane cannot continue slipping turn, because there are no more aileron deflections available to maintain the 30° bank angle. Instead, the airplane increases speed, which consequently increases aileron effectiveness and yields in less aileron deflection requirement for the same amount of turn rate. In the other words, the airplane now can access higher turn rates by increasing speed, with the same saturated aileron. Hence, it yaws more to the left (increases sideslip) via increasing positive rudder input, while the aileron and bank angle are saturated at 20° and 30°, respectively. It is important to note that aileron saturation is the limiting factor, whilst bank angle saturation is the constraint being satisfied, not the limiting factor. The airplane inputs more throttle, increases speed, and achieves higher turn rates until engines' thrust is saturated (end of the aileron saturation boundary, beginning of the thrust saturation boundary). From this point onwards, the airplane goes on with full throttle. Since engines' available thrust is decreasing while increasing speed; the airplane decreases required thrust by reducing drag through decreasing sideslip via reducing rudder deflection angle. Reduction of sideslip angle decreases airplane's reference area ($S$) in $D = \frac{1}{2}\rho V^2 S C_D$, where $D$ is drag force. In the red colored section of the boundary, this reference area reduction is such that enables the airplane to increase speed while total drag force (and hence, the required thrust) is decreasing. The airplane's turn rate reduces as it decreases sideslip angle. It keeps increasing speed and reducing sideslip; until it enters coordinated turn (zero sideslip), and from that point, bank angle decreases until the airplane is in non-sideslipping wings level rectilinear flight at top of the flight envelope (end of the thrust saturation boundary).

Stall speed increases as the aircraft increases bank angle:

$$L/W = 1/\cos\phi = n \qquad (28)$$

$$V_S = \sqrt{2L/\rho C_{L_{max}} S} = \sqrt{2nW/\rho C_{L_{max}} S} \qquad (29)$$

$$V_{S_t} = \sqrt{n} V_{S_0} = 1/\sqrt{\cos\phi} V_{S_0} \qquad (30)$$

where, $n, W, V_{S_t}, V_{S_0}$ are load factor, weight, turning maneuver stall speed, and wings level rectilinear flight stall speed. Stall speed is proportional to load factor as presented in (29), (30), and load factor increases with roll angle; as shown in (28). Stall speed increase can be seen in the stall boundary section of figures 5 and 8. However, instead of being gradual, such increase is in the form of $V$-constant sections.

$V$-constant sections in the stall boundary of the unimpaired flight envelope shown in Fig. 5, are generally due to numeric round-off error and resolution increment size of $V$ range (1 knot). In fact, the speed increases gradually (according to (30)), and if the increment size is reduced significantly (e.g. to 0.1 knots), these stepwise constant sections would be eliminated and replaced by a curve of boundary trim points. However, this would exponentially increase the computational cost which is unnecessary, because we already have estimated sufficiently accurate flight envelope boundary, with current increment size.

The effect of these $V$-constant sections can be seen as "sawtooths" in the zoomed-in-plot of figure 6. In Fig. 6, these sawtooths can be described as increase in the angle of attack in $V$-constant sections, and sudden decrease in the angle of attack between the $V$-constant sections. As the aircraft banks, lift vector inclines, hence there is lesser lift available to balance the weight. Therefore, the angle of attack must be increased to prevent loss of altitude. This is the reason for the increase of angle of attack in $V$-constant sections. In each $V$-constant section, the angle of attack increases until it reaches the critical value. So, the extra lift required to prevent loss of altitude cannot be provided by increasing angle of attack anymore. At this point, the required lift is generated by increasing speed, and the current $V$-constant section changes to the next $V$-constant section. Between the $V$-constant sections; where extra required lift is provided by an increase in the speed, there would be lesser need in the angle of attack, and that is why the angle of attack decreases between the $V$-constant sections.

$\phi$-constant and $\beta$-constant thrust-required curves are plotted in Fig. 7. Usually in texts and articles, only $\phi = 0$ thrust-required curve is plotted along with thrust-available curve. However in this research, we have additionally calculated and plotted "$\phi = 30°$, $\beta$-constant" curves for various amounts of sideslip angle to show that as we move along the trim points of the flight envelope boundary, the required thrust increases and boundary trim points pass through $\beta$-constant curves. For better realization, these curves are shown in the unimpaired flight envelope boundary in Fig. 12. These curves are confined by the aileron saturation boundary on one side and the thrust saturation boundary on the other side.

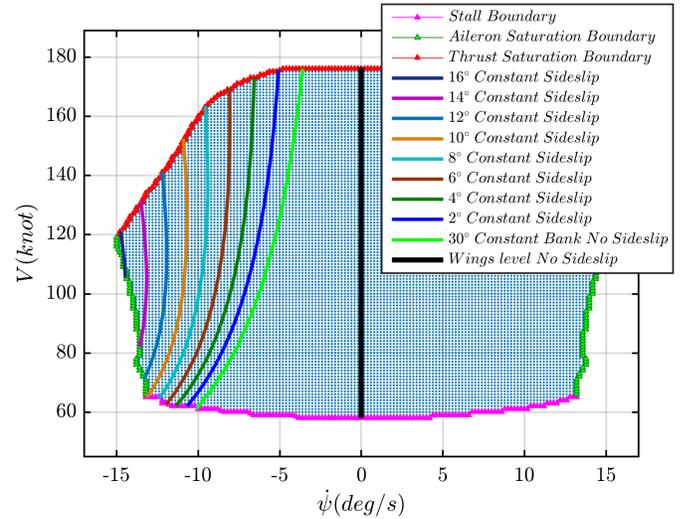

Fig. 12. $\beta$-constant curves in unimpaired flight envelope at sea level and zero flight path angle

There are segments in the green colored boundary of Fig. 10, where one trim point has saturated aileron of 20°, whilst in the next few trim points, the trim value of aileron deflection is slightly lower than 20°. In these segments, speed increases with constant turn rate, which increases the aileron effectiveness, and hence lesser aileron (< 20°) is needed for the same amount of turn rate. However, this does not mean that the airplane can access the adjacent trim point if it uses full 20° aileron deflection. This is because we have $\dot\psi$ increments of 0.2 deg/s and the adjacent trim point ($\dot\psi + 0.2$) requires aileron deflection of greater than 20°.

Similar behavior is seen in the thrust saturation boundary of Fig. 7. There are segments in which a trim point has 100% throttle setting whilst the next few ones do not fully lie on the thrust-available curve. Again, the reason is the increment size of turn rate. At such trim points, if the turn rate is increased by 0.2 deg/s, the required thrust would be more than the available thrust, and if we wish to stay on the thrust-available curve, we need to increase the turn rate by less than 0.2 deg/s, which is not available. However, after increasing speed by three or four knots, the next trim point lies on the thrust-available curve again, which is resulted by simultaneous increase in speed and decrease in available thrust. A sample segment is chosen in Fig. 7 to demonstrate this behavior. Zoomed-in-plot of this segment shows five trim points which the first and the last ones are placed on the thrust-available curve, whereas the other three are beneath the curve. As shown in Table 6, these three trim points require greater than available amounts of thrust to become their adjacent trim points (same $V$, $\dot{\psi} + 0.2$).

TABLE 6
THRUST REQUIRED FOR THREE SAMPLE TRIM POINTS

| Speed | Required Throttle Setting |
|---|---|
| 135 knot | 102.09% |
| 136 knot | 102.30% |
| 137 knot | 100.07% |

Despite the fact that left aileron (positive $\delta_a$) is needed to initiate a roll to the left, it is shown in figures 10 and 11 that right aileron (negative $\delta_a$) and left rudder (positive $\delta_r$) are required to perform a left turn (negative $\phi$) maneuver. The reason that right aileron is needed to maintain the bank angle constant during the left coordinated turn, can be explained as following: As the airplane banks to the left, an adverse yawing moment is generated which yaws the airplane to the right while the airplane is rolling to the left. To counteract this yawing moment, some left rudder input is required, which its exact amount is primarily dependent on yaw damping derivative of the airplane. Due to roll-yaw coupling, this rudder input produces some right increment in the rolling moment; however, it is not enough to compensate for the effects of flight path curvature. Hence, some right aileron input is also required [44].

Unimpaired flight envelopes of GTM at four different altitudes of sea level, 10000 ft, 20000 ft, and 30000 ft are presented in Fig. 13. It can be seen that flight envelope contracts as the altitude increases. To understand why flight envelope shrinks, we need to investigate the boundaries.

According to (29), it is evident that by increasing altitude; stall speed increases too, which is due to decrease in the air density. Hence, the magenta colored segment of the boundary (the stall boundary) goes up with the increase in altitude.

As shown in Fig. 14, increasing altitude results in reduction of engines' available thrust. This leads to lesser maximum speed ($V_{max}$) at the higher altitude. This is why the red colored segment of the boundary (thrust saturation boundary), contracts and lowers with the increase in altitude.

As the altitude increases, the aileron saturation boundary (green colored segment) gets smaller and the aileron saturates sooner (at smaller turn rates), until 30000 ft where the aileron saturation boundary completely vanishes. To understand why this happens, the result of an investigation done on the intersection trim point of stall boundary and aileron saturation boundary; at different altitudes, is presented in Table 7.

TABLE 7
INTERSECTION TRIM POINT STATUS AT VARIOUS ALTITUDES

| Altitude | Aileron Deflection | Throttle Setting |
|---|---|---|
| Sea level | Saturated | 45.4% |
| 10000 ft | Saturated | 62.9% |
| 20000 ft | Saturated | 88.3% |
| 30000 ft | -13.36° | 98.7% |

Based on Table 7, at sea level the aileron is saturated whilst thrust is far from saturation. However, by increasing altitude, the point where the stall boundary ends and the aileron saturates will have lesser distance from thrust saturation. Therefore, the green segment of the boundary becomes smaller. As the aircraft arrives at 30000 ft, there is no distance left between the thrust saturation boundary and the end of the stall boundary. So, thrust is saturated long before the aileron saturates ($\delta_a = -13.36°$) and the green segment disappears.

Also due to decrease in the air density, aileron effectiveness reduces with increase in the altitude. Hence more rudder input is required to achieve the same turn rate of the previous altitude, which means more aileron deflection is needed to maintain the bank angle constant at 30°, and consequently the aileron saturates sooner.

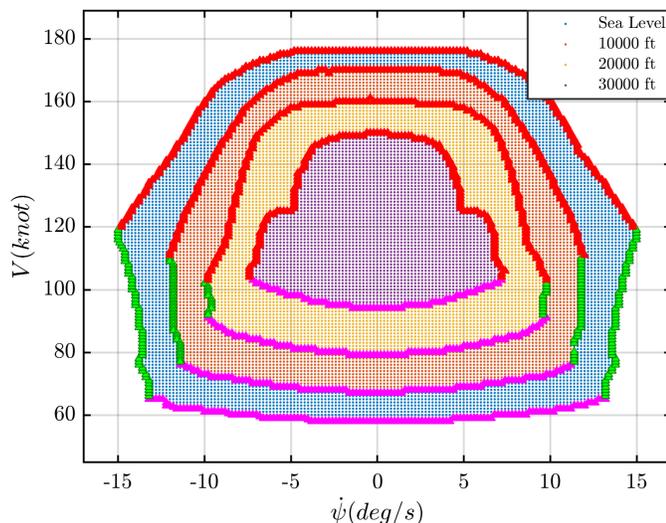

Fig. 13. Unimpaired flight envelopes at zero flight path angle and various altitudes

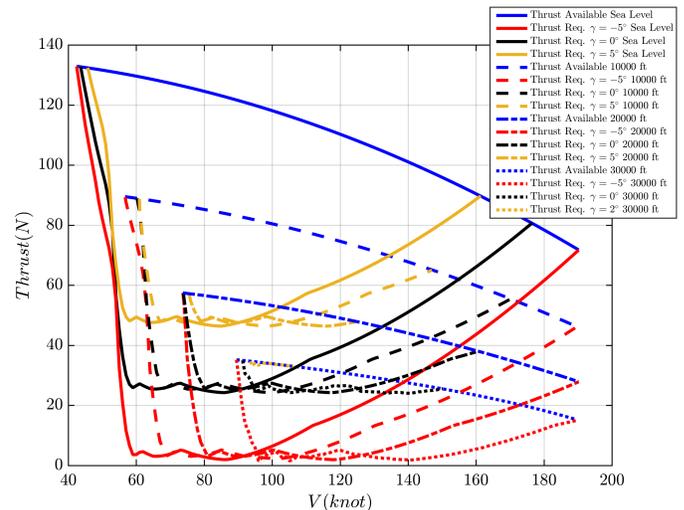

Fig. 14. Thrust-available – Thrust-required curves for various flight path angles and different altitudes

As seen in Fig. 13, there is a sudden indent on each side of the thrust saturation boundary of the 30000 ft flight envelope. Source of these indents is in the thrust-required curve of the GTM. There are few slight ups and downs at the bottom of all thrust-required curves of the GTM (see Fig. 14), which are due to the aerodynamic coefficients in the aero datasets. At 30000 ft, the distance between thrust-required curve and thrust-available curve is such limited that for some trim points with speeds corresponding to the mentioned ups, the amount of the required thrust for many turn rates fall beyond the thrust-available curve. Hence, to be able to trim the aircraft at mentioned speeds, the turn rate needs to be reduced significantly.

Fig. 15 shows unimpaired flight envelopes for various flight path angles at sea level. As expected, the stall boundary and the aileron saturation boundary are almost at the same location in different flight path angles. This is due to the fact that flight path angle has no significant effect on stall speed value or saturation of aileron.

However, thrust saturation boundary shifts down by increasing flight path angle, which means lesser thrust and lesser maximum speed ($V_{max}$) are available as the flight path angle increases. This is completely in compliance with the following equation:

$$T_R = W\gamma + \frac{1}{2}\rho V^2 S C_{D_0} + 2KW^2 / \rho V^2 S cos^2\phi \qquad (31)$$

in which, $T_R$ and $K$ are required thrust and drag polar parameter, respectively [45]. According to (31), the required thrust increases with increase in flight path angle. This is also evident in Fig. 14. On the other hand, thrust available curve is the same for different flight path angles, because they are all considered at the same altitude. Consequently, $V_{max}$ reduces as flight path angle increases.

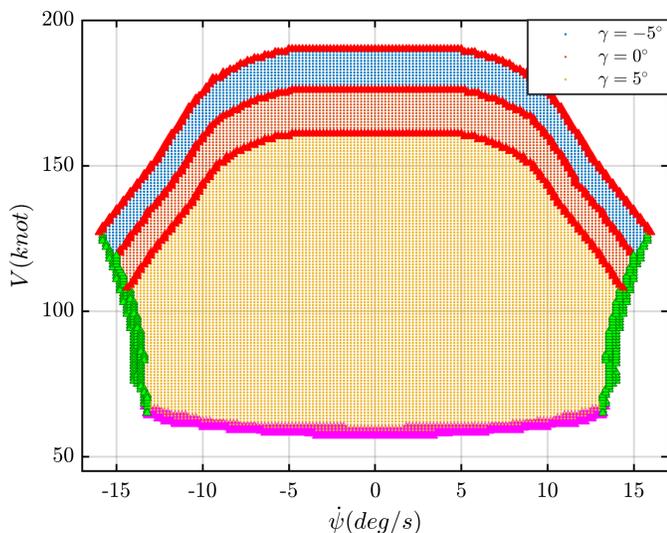

Fig. 15. Unimpaired flight envelopes at sea level and various flight path angles

It is important to note that at 30000 ft, the airplane can fly only up to 2° flight path angle. In the other words, there are no climbing steady maneuvers feasible for the GTM with $\gamma > 2°$ at 30000 ft. This is because the thrust-available curve has dropped significantly and there is very limited thrust available at this altitude. Also, thrust-required curve has elevated with each incremental increase of flight path angle, as in (31). For flight path angles greater than 2°, all required thrust values are higher than all available thrust values.

The maximum available flight path angle can be computed from the following equation:

$$\gamma_{max} = \frac{T_{max}^S}{W}\left(\frac{\rho}{\rho^S}\right)^m - 2cos\phi\sqrt{KC_{D_0}} \qquad (32)$$

where $T_{max}^S$ and $\rho^S$ are maximum engines' thrust and air density at sea level. Using (32) at 30000 ft also results in 2° maximum flight path angle.

By the way, it should be noted that the GTM flies at lower altitudes than the full-scale aircraft [46]. So the 2° limit of flight path angle does not exist for the full-scale aircraft at 30000 ft, and it can achieve higher flight path angles at this altitude.

### 5.2. Impaired flight envelopes

In this section we aim to investigate the variation of flight envelope boundary with different degrees of control surface failure. Hence, impaired flight envelopes at a number of flight conditions are plotted for comparison purposes.

#### 5.2.1. Rudder restriction cases

By analyzing the boundaries of the impaired flight envelopes shown in figures 16 to 18, we can conclude that in the rudder restriction cases, the right side of the flight envelope boundary corresponds to the lower limit of the rudder deflection angle, whereas the left side of the boundary corresponds to the upper limit. The reason is that right deflection of rudder (negative $\delta_r$) causes positive yawing moment (and its corresponding adverse roll), which initiates turn to the right (positive turn rates). On the other hand, when rudder deflects to the left (positive $\delta_r$), the generated negative yawing moment results in negative turn rates (turn to the left).

As shown in the presented figures, when rudder deflection is restricted, flight envelope contracts and the impaired boundary separates from the boundary of the unimpaired case.

This separation occurs at that side of the flight envelope which its corresponding limit is lower than that of the unimpaired case (e.g. the limit being 10° instead of 30°). For instance, in Fig. 16, all of the three impaired cases [−30, 20], [−30,10], and [−30,0], have right boundaries attached to the right boundary of the unimpaired flight envelope. Because their lower limit of the rudder deflection angle is the same as the lower limit in the unimpaired aircraft (i.e. −30°). However, their upper limits are lower than 30°, so their left boundaries are separated from the left boundary of the unimpaired case, where the amount of separation (distance between their left boundary and the unimpaired boundary) is proportional to the value of their upper limit. As the upper limit value lowers, the failure gets severe (i.e. control restriction tightens), so the impaired boundary retreats more inside the unimpaired flight envelope and distances more from the unimpaired boundary.

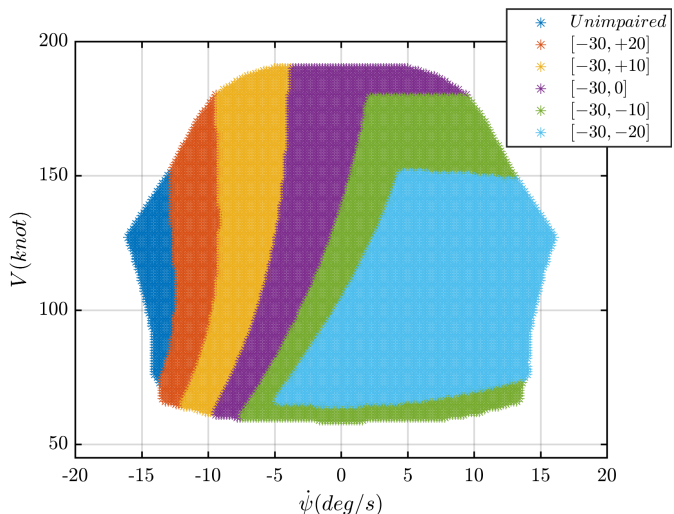

Fig. 16. Rudder restriction cases at sea level and $\gamma = -5°$

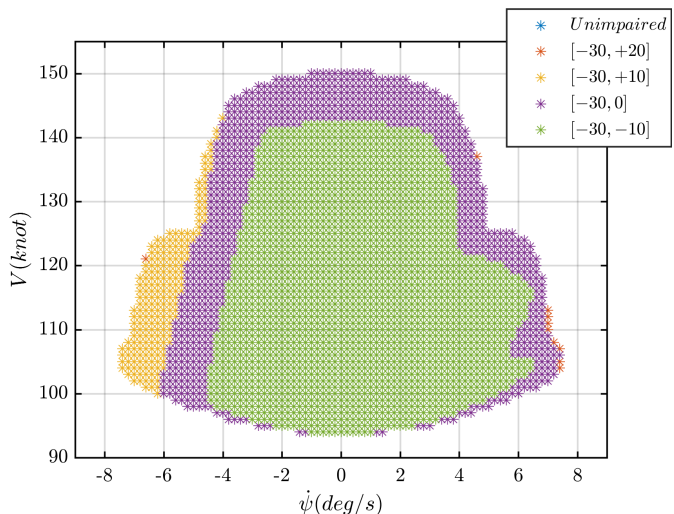

Fig. 17. Rudder restriction cases at 30000 ft and $\gamma = 0°$

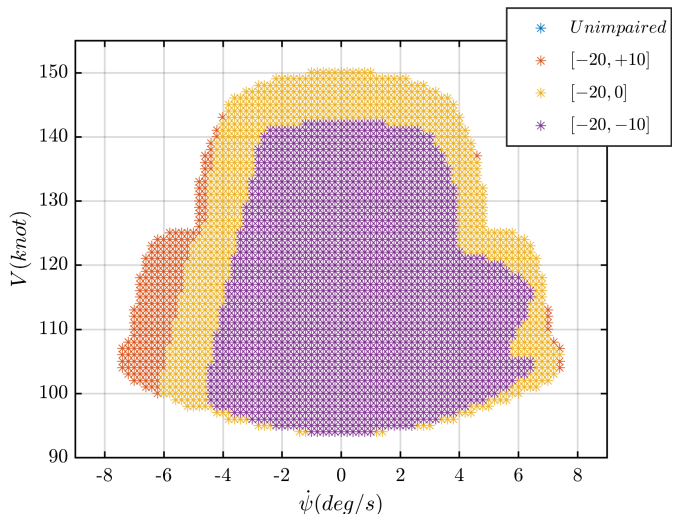

Fig. 18. Rudder restriction cases at 30000 ft and $\gamma = 0°$

Another outcome of the performed analysis is shown in Table 8 and Table 9 in which ✓ represents the existence of difference between the corresponding impaired boundary and the unimpaired boundary, whereas ✗ means no difference.

TABLE 8
[−20, UL] / [LL, 20] DIFFERENCE WITH UNIMPAIRED BOUNDARY

| Altitude | $\gamma = -5$ | $\gamma = 0$ | $\gamma = 5$ |
|---|---|---|---|
| Sea level | ✓ | ✓ | ✓ |
| 10000 ft | ✓ | ✓ | ✗ |
| 20000 ft | ✓ | ✗ | ✗ |
| 30000 ft | ✗ | ✗ | — |

TABLE 9
[−10, UL] / [LL, 10] DIFFERENCE WITH UNIMPAIRED BOUNDARY

| Altitude | $\gamma = -5$ | $\gamma = 0$ | $\gamma = 5$ |
|---|---|---|---|
| Sea level | ✓ | ✓ | ✓ |
| 10000 ft | ✓ | ✓ | ✓ |
| 20000 ft | ✓ | ✓ | ✗ |
| 30000 ft | ✓ | ✗ | — |

Table 8 shows at which altitude and flight path angle, the right boundary of the impaired flight envelope in [−20, UL] case (*or the left boundary in [LL, 20] cas*e) is the same as the right (*or left*) boundary of the unimpaired flight envelope, and at which altitude and flight path angle, it is separated.

According to Table 8, as flight path angle and altitude increase, it is more likely that the unimpaired and impaired cases have the same corresponding boundary. That is because by increasing altitude or flight path angle, the unimpaired flight envelope contracts and thrust saturation boundary becomes more dominant (see Fig. 13 and Fig. 15). This contraction and domination leads to smaller amounts of required rudder deflection angle at the boundaries of the unimpaired flight envelope, because further trim points are infeasible due to lack of thrust. So when the limit of the restricted rudder is bigger than the required deflection values of the unimpaired boundary trim points, the impaired aircraft can be trimmed at the unimpaired boundary trim points even with the restricted rudder. Similar explanation is applicable to Table 9.

For example, at sea level and zero flight path angle, the maximum rudder deflection angle used in unimpaired boundary trim points is ±26° (Fig. 11), however this value reduces to ±7° at 30000 ft because trim maneuvers which require rudder deflection of more than ±7° are infeasible due to thrust saturation. Therefore any rudder restriction failure in which the restricted lower or upper limit is beyond ±7°, impose no restriction on airplane's maneuverability at 30000 ft (Table 9).

In Fig. 17, flight envelope of the impaired case [−30, 10] is the same as the unimpaired case [−30, 30], because the furthest trim point on the boundary of the unimpaired flight envelope requires less than 10° rudder deflection, hence the restriction has no effect (Table 9). This is also the reason that the unimpaired flight envelope is not visible in that figure. It is completely covered by the impaired flight envelopes [−30, 10] and [−30, 20].

The impaired case [−20, 10] in Fig. 18 corresponds to both of the tables 8 and 9. Neither the lower limit boundary nor the upper limit boundary has any difference with the unimpaired flight envelope boundary.

It should be noted that the symmetry of the estimated flight envelopes (described in the beginning of this section), results in validity of Table 8 for both [−20, UL] and [LL, 20] cases. With the same reason, Table 9 is valid for both [−10, UL] and [LL, 10] cases.

It can be seen that in most cases, top (thrust saturation boundary) and bottom (stall boundary) of the impaired flight envelope are attached to top and bottom of the unimpaired flight envelope, except for that side of the boundary which is separated. In such cases, the deflection range of the impaired rudder includes 0° deflection, which means that the aircraft is able to perform non-sideslipping maneuver ($\beta == 0$). However, there are cases in which impaired rudder deflection range excludes 0° deflection. In these cases there are either positive rudder inputs available or negative rudder inputs available. Thus the aircraft is always sideslipped ($\beta \neq 0$), which yields in persistent extra drag force. Such drag increases required thrust, which in turn leads to lower maximum speed and higher minimum speed. We call these cases "high drag". It is evident in the presented figures that for the high drag cases, always top of the impaired boundary and sometimes (if the restriction is highly tightened) bottom of the impaired boundary separates from the unimpaired boundary. For instance in Fig. 16, [−30, −10] and [−30, −20] are both high drag cases.

Figures 16, 17, and 18 are three instances among numerous comparison plots. More comparison plots are provided in [41].

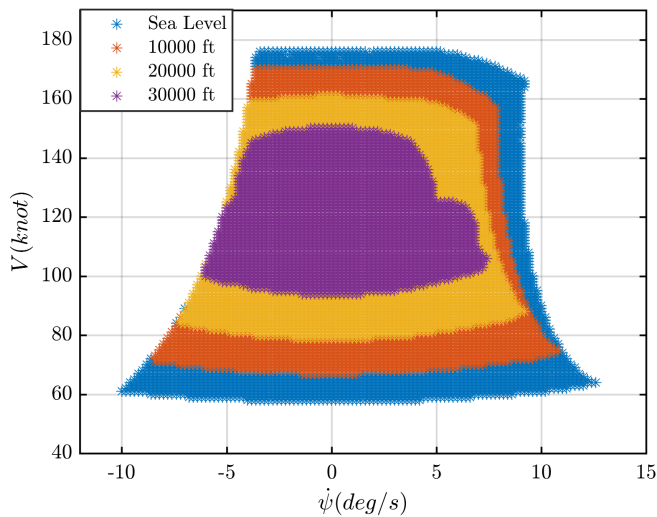

Fig. 19. Rudder restriction [−10, 0] at zero flight path angle

It is worthy to check the validity of the results of the previous unimpaired flight envelope analysis in the impaired cases too. Therefore, three impaired cases are chosen and plotted in different altitudes and flight path angles (figures 19, 20, and 21). In Fig. 19, impaired flight envelopes shrink as expected with the increase in altitude, which is consistent with Fig. 16. Fig. 20 demonstrates lowering of the thrust saturation boundary with increase in the flight path angle, which is due to increase in the required thrust. Same behavior is seen in Fig. 15 for the unimpaired cases. In Fig. 21, despite the contraction of the impaired flight envelope, maximum speed is the same for different altitudes. A phenomenon completely in compliance with Fig. 17, which shows that in $\gamma = -5°$, maximum speed does not change with the variation in altitude.

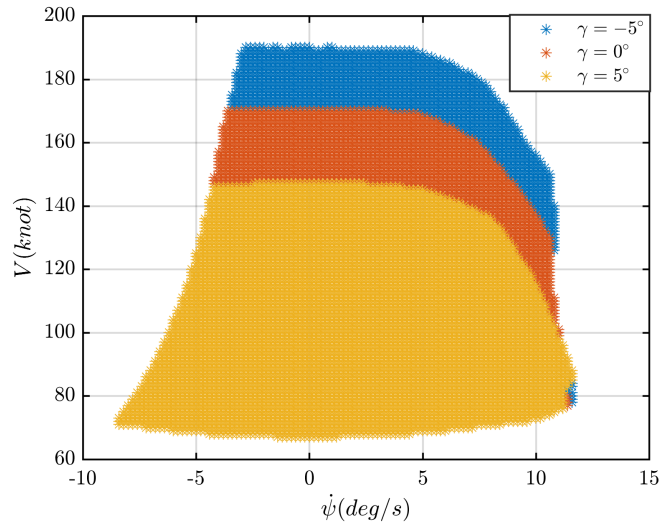

Fig. 20. Rudder restriction [−20, 0] at 10000 ft

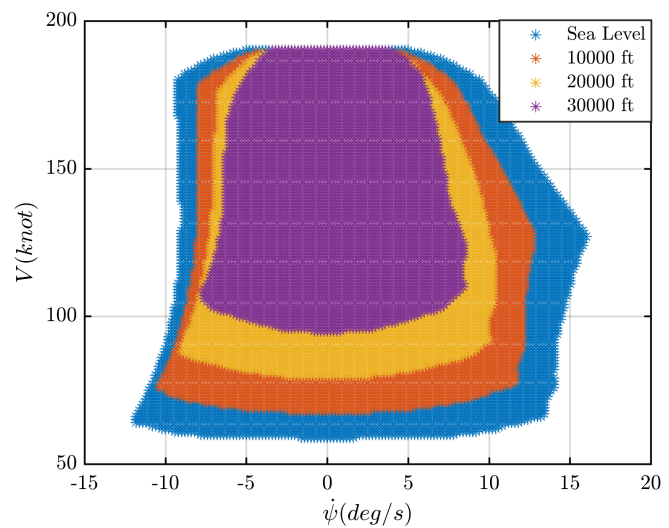

Fig. 21. Rudder restriction [−30, 10] at $\gamma = -5°$

### 5.2.2. Aileron restriction cases

General rules inferred from the rudder restriction cases are valid for the aileron restriction cases too, except for few details changed. Lower limit of the aileron deflection angle corresponds to the left side of the flight envelope boundary, because right aileron (negative $\delta_a$) incurs a positive rolling moment which is followed by an adverse negative yawing moment which makes left turns (negative turn rates) more accessible. Consequently, upper limit corresponds to the right side of the flight envelope boundary.

As in the rudder restriction cases, restricted aileron failures which do not include the 0° deflection angle are always sideslipped ($\beta \neq 0$) and endure permanent extra drag force. For example, the impaired case [−20, −10] in Fig. 22 is a high drag case.

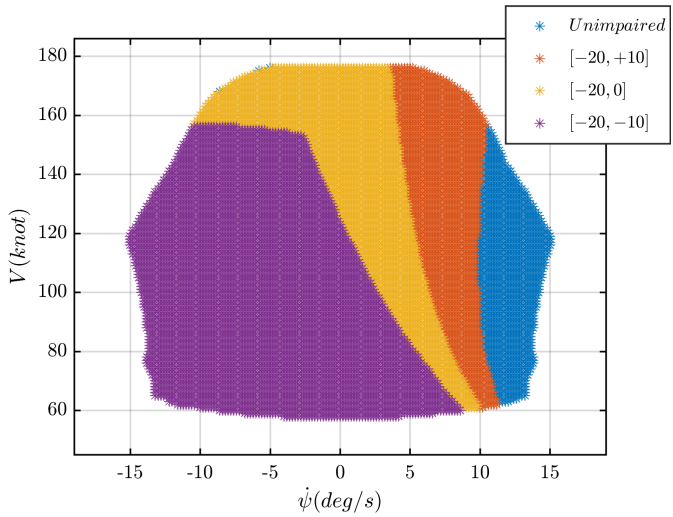

Fig. 22. Aileron restriction cases at sea level and $\gamma = 0°$

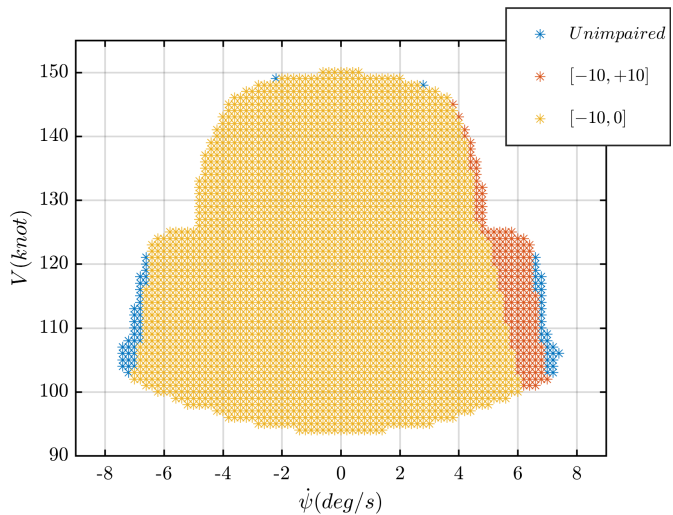

Fig. 23. Aileron restriction cases at 30000 ft and $\gamma = 0°$

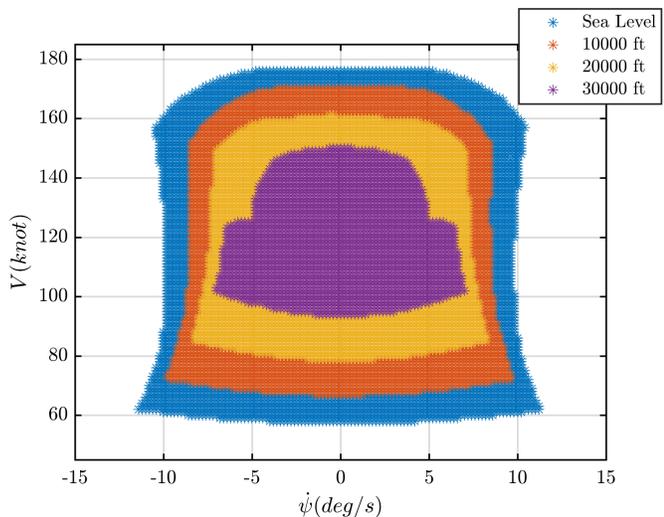

Fig. 24. Aileron restriction [−10, 10] at $\gamma = 0°$

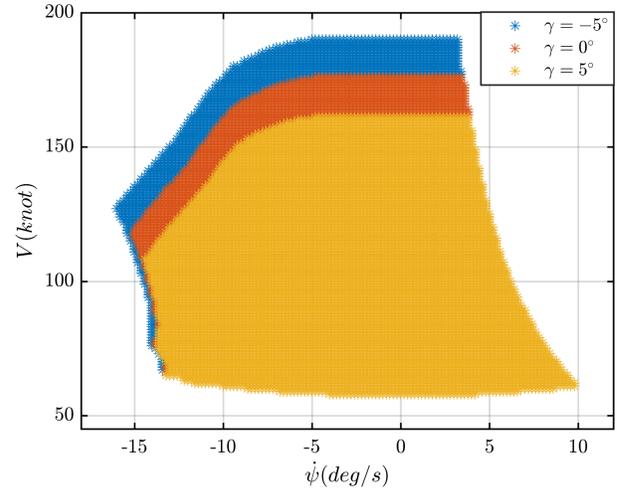

Fig. 25. Aileron restriction [−20, 0] at sea level

In contrast to the rudder restriction cases, there is no impaired aileron flight envelope boundary that is the same as the unimpaired boundary. For instance in Fig. 23, the right side of the boundary of aileron impaired case [−10, 10] and the left side of the boundary of the impaired case [−10, 0] are very close to the corresponding unimpaired boundary, however they have very slight differences. This means that in all aileron restriction failures, the effect of the unavailability of part of the aileron deflection range is more dominant than the thrust saturation.

Eventually, flight envelope variation of the impaired case [−10, 10] with the increase in altitude, and flight envelope variation of the impaired case [−20, 0] with the increase in flight path angle are demonstrated in figures 24 and 25, respectively. As expected, the variations are consistent with figures 13 and 15.

### 5.2.3. Rudder jamming cases

Figures 26 and 27 present maneuvering flight envelopes for a number of rudder jamming failure cases. These figures show that when the rudder is jammed at a positive deflection angle, the aircraft tends to turn left. That is because positive rudder deflection (left rudder) incurs negative yawing moment. Also it can be seen that as the rudder jammed angle increases (failure gets severe), flight envelopes are more drifted to the left, i.e. left turning maneuvers with higher turn rates become accessible whilst turning maneuvers with lower turn rates become infeasible. This behavior can be explained as following:

In the unimpaired case, when the bank angle reaches its maximum value (according to the imposed 30° bank constraint), the airplane can still achieve higher turn rates by performing slipping turn. To do so, simultaneous opposite rudder and aileron inputs are used to yaw the aircraft whilst maintaining the bank angle at 30°. However, in the impaired cases there are no rudder inputs available when the aircraft reaches maximum allowable bank angle (because the rudder is jammed), so it is impossible for the airplane to perform slipping turn and maintain bank angle. Thus further turn rates and trim points are infeasible.

Also it is evident that the rudder jammed airplane has permanent sideslip ($\beta \neq 0$), which yields in shifting the flight envelopes of the impaired cases to the left (more negative turn rates) in the $\phi - \dot{\psi}$ plot. The more the rudder jammed angle, the more the shifting, and hence the more drifted is the impaired flight envelope.

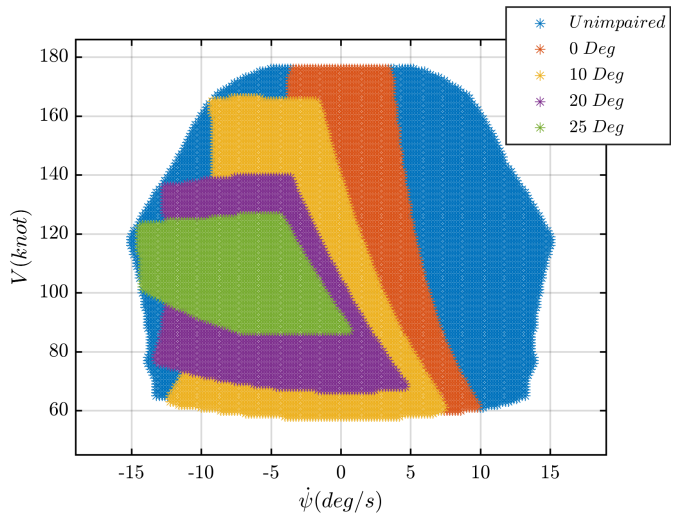

Fig. 26. Rudder jamming cases at sea level and $\gamma = 0°$

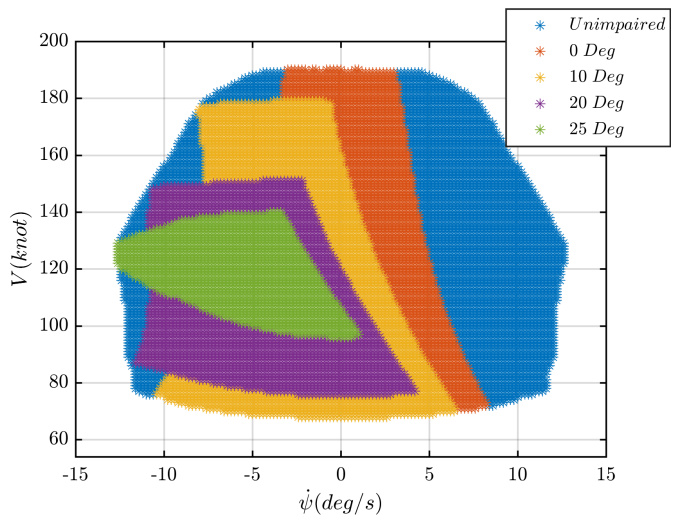

Fig. 27. Rudder jamming cases at 10000 ft and $\gamma = -5°$

By investigating the maneuvering flight envelopes of the rudder jamming cases and comparing them with the restricted rudder flight envelopes, we arrive at the following conclusion:

Each jammed rudder flight envelope is the intersection of two restricted rudder flight envelopes. This statement can be shown as below:

$$J[X] = R[LL, X] \cap R[X, UL] \tag{33}$$

where $J$ denotes Jamming, $R$ denotes restriction, and $X$ is the jammed rudder angle. For instance, maneuvering flight envelope of aircraft with jammed rudder at **+20°** is comprised of trim points which are available in both of the flight envelopes of the restricted rudder cases [−30°, **+20°**] and [**+20°**, +30°]. This is shown in Fig. 28. Also in Fig. 29, it is presented that the maneuvering flight envelope of aircraft with jammed rudder at **+10°** is the intersection of flight envelopes of restriction cases [−20°, **+10°**] and [**+10°**, +20°].

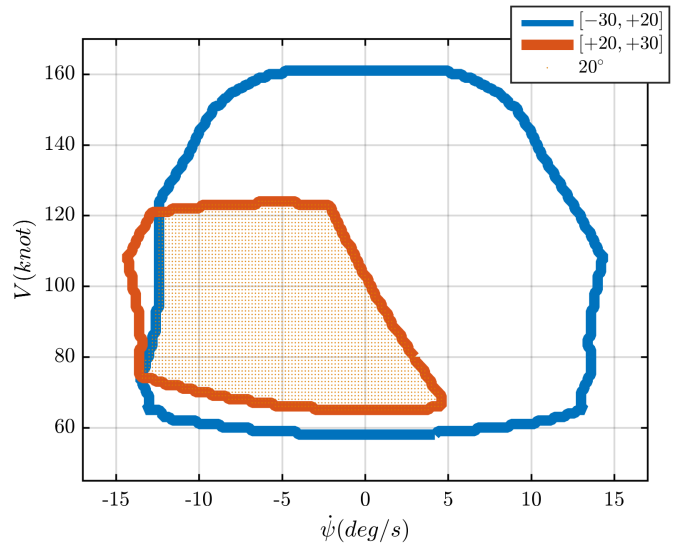

Fig. 28. Intersection of flight envelopes at sea level and $\gamma = 5°$

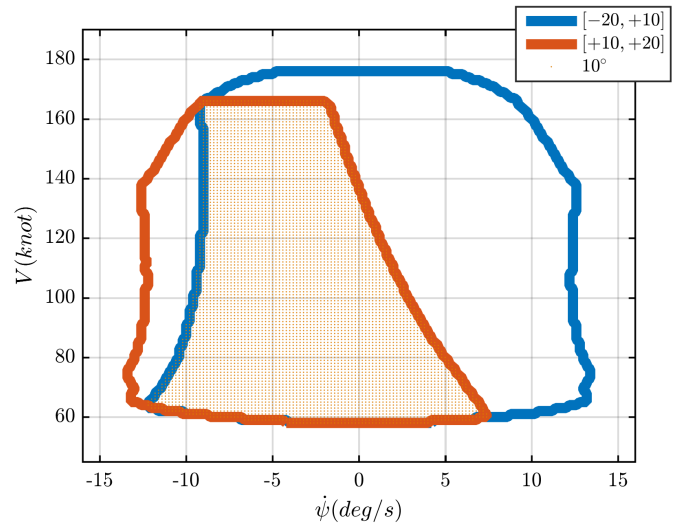

Fig. 29. Intersection of flight envelopes at sea level and $\gamma = 0°$

### 5.2.4. Aileron jamming cases

Figures 30 and 31 present maneuvering flight envelopes for a number of aileron jamming failure cases. As can be seen, when aileron is jammed at a positive value, the aircraft tends to turn to the right. That is because left aileron (positive $\delta_a$) incurs a negative rolling moment which is followed by an adverse positive yawing moment which makes right turns (positive turn rates) more accessible.

As the aileron jammed angle increases, maneuvering flight envelopes are more drifted to the right, so higher turn rates become feasible whilst lower turn rates become inaccessible.

Similar to the intersection rule inferred from the rudder jamming failures exists for the aileron jamming cases. Figure 32 demonstrates an instance.

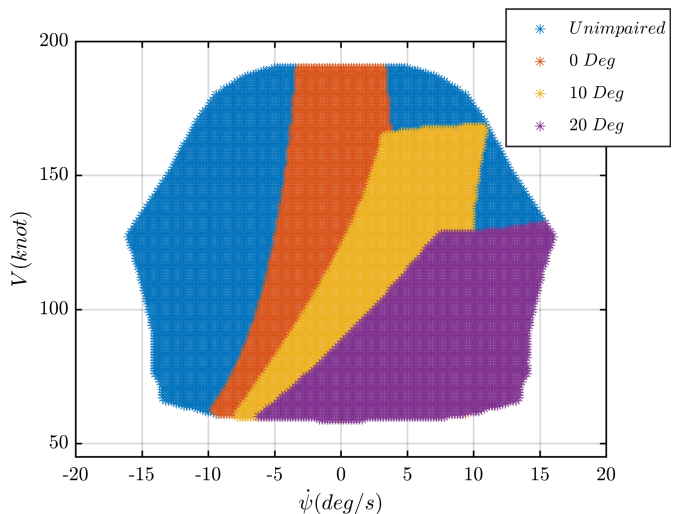

Fig. 30. Aileron jamming cases at sea level and $\gamma = -5°$

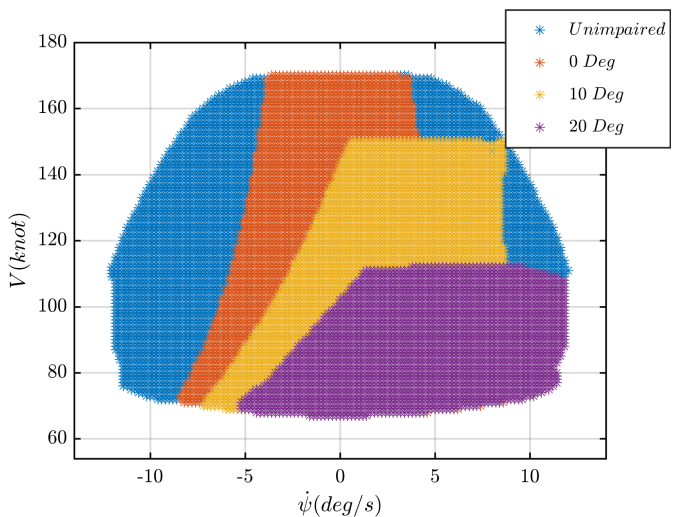

Fig. 31. Aileron jamming cases at 10000 ft and $\gamma = 0°$

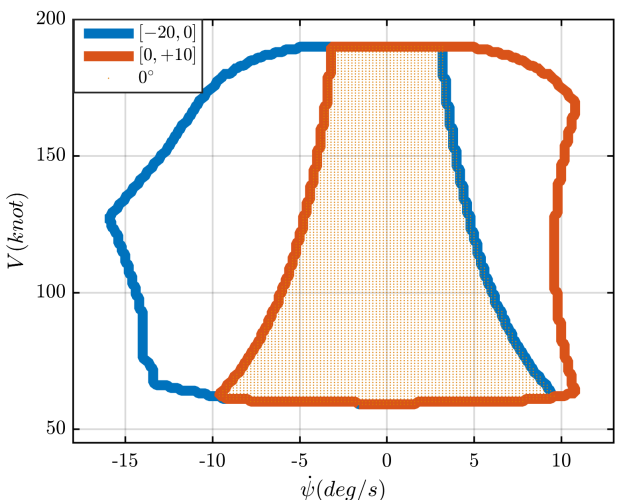

Fig. 32. Intersection of flight envelopes at sea level and $\gamma = -5°$

## 6. Conclusion

In this paper, the effect of rudder and aileron failures on the maneuvering flight envelope of the NASA GTM has been investigated. To do so, first a rigorous analysis was performed on the maneuvering flight envelope of the unimpaired GTM to identify the limiting factors of the flight envelope boundary. Precise cognition of the unimpaired flight envelope is the key prerequisite to understand the variation of the maneuvering flight envelope in failure cases. Hence, unlike the other related researches in literature where often no detailed study has been done on the unimpaired maneuvering flight envelope and directly brief results of the impaired flight envelopes are provided, in this study we have thoroughly investigated the flight envelope of the unimpaired GTM in various flight conditions. The unimpaired flight envelope was evaluated at different altitudes and flight path angles, and the states and controls variation along the boundary trim points were depicted and analyzed. Results show that the flight envelope contracts as the altitude increases, because the thrust saturation boundary contracts and lowers, the stall boundary (stall speed) goes up, and the aileron saturation boundary gets smaller (the aileron saturates sooner) with the increase in altitude. Also, the stall boundary and the aileron saturation boundary are almost at the same location at different flight path angles; however, thrust saturation boundary shifts down by increasing the flight path angle. The obtained results were also verified by the corresponding equations governing the aircraft performance.

In this research, an extensive previously created database comprised of 164 3D maneuvering flight envelopes of 41 different surface jam and control restriction failures of the GTM's rudder and aileron at different flying altitudes was employed. Each 3D flight envelope contains thousands of trim points for which a nonlinear 6DOF optimization process has been executed one by one. On average it took 8 seconds for each trim point to complete the optimization process. Hence, in order to evaluate all the 3D flight envelopes of the database, more than 19600 hours was spent in over more than 16 months; executing the optimization process on more than 8.8 million trim points. Such a large number of impaired flight envelopes provided a great insight on how the degree of failure affects the maneuvering flight envelope. Specifically, results show that in both of the rudder and aileron failure cases, flight envelope contracts and the impaired boundary separates from the boundary of the unimpaired case, and this separation occurs at that side of the impaired flight envelope which its corresponding limit is lower than that of the unimpaired case (e.g. the limit being 10° instead of 30°). As the upper limit value of the control surface deflection decreases or the lower limit value increases, the failure gets severe (i.e. control restriction tightens), so the impaired boundary retreats more inside the unimpaired flight envelope and distances more from the unimpaired boundary. Also, there are failure cases in which, impaired deflection range excludes the 0° deflection. In these cases, there are either positive deflection inputs available or negative inputs available. Thus, the aircraft is always sideslipped ($\beta \neq 0$), which yields in persistent extra drag force. Such drag increases the required thrust, which in turn leads to lower maximum speed and higher minimum speed. Hence, such failure cases have more shrunk flight envelopes. Also, results of the study show that each jamming case's flight envelope is the intersection of two restriction cases' flight envelopes.

Future works include the evaluation and analysis of the impaired flight envelopes for other types of failure, such as structural damages, engine malfunctions, and icing. The effect of the combination of different failure types is also an interesting topic.

## CONFLICTS OF INTEREST

The authors declare that there is no conflict of interest regarding the publication of this paper.

## FUNDING STATEMENT

This research did not receive any specific grant from funding agencies in the public, commercial, or not-for-profit sectors.

## REFERENCES

[1] "Statistical Summary of Commercial Jet Airplane Accidents Worldwide Operations | 1959 – 2017," Aviation Safety, Boeing Commercial Airplanes, Seattle, WA, Oct. 2018.

[2] "Global Fatal Accident Review 2002 - 2011," TSO (The Stationery Office) on behalf of the UK Civil Aviation Authority, Norwich, UK, 2013 [Online]. Available: http://publicapps.caa.co.uk/cap1036. [Accessed: 02- May- 2017].

[3] Gill SJ, Lowenberg MH, Neild SA, Krauskopf B, Puyou G, and Coetzee E, "Upset Dynamics of an Airliner Model: A Nonlinear Bifurcation Analysis," *Journal of Aircraft*, vol. 50, no. 6, pp. 1832–1842, Nov. 2013. DOI: 10.2514/1.C032221, [Online].

[4] Norouzi, R., Kosari, A., and Sabour, M. H., "Investigating the Generalization Capability and Performance of Neural Networks and Neuro-Fuzzy Systems for Nonlinear Dynamics Modeling of Impaired Aircraft," *IEEE Access*, vol. 7, pp. 21067–21093, Feb. 2019. DOI: 10.1109/ACCESS.2019.2897487, [Online].

[5] Norouzi, R., Kosari, A., Sabour, M.H., "Real Time Estimation of Impaired Aircraft Flight Envelope using Feedforward Neural Networks," *Aerospace Science and Technology*, May 2019. DOI: 10.1016/j.ast.2019.04.048, [Online].

[6] Wilborn J, and Foster J, "Defining Commercial Transport Loss-of-Control: A Quantitative Approach," in *AIAA Atmospheric Flight Mechanics Conference and Exhibit*, 2004 [Online]. Available: http://dx.doi.org/10.2514/6.2004-4811

[7] Yi G, and Atkins E, "Trim State Discovery for an Adaptive Flight Planner," in *48th AIAA Aerospace Sciences Meeting Including the New Horizons Forum and Aerospace Exposition*, 2010 [Online]. Available: http://dx.doi.org/10.2514/6.2010-416

[8] Vannelli A, and Vidyasagar M, "Maximal lyapunov functions and domains of attraction for autonomous nonlinear systems," *Automatica*, vol. 21, no. 1, pp. 69–80, Jan. 1985. DOI: 10.1016/0005-1098(85)90099-8, [Online].

[9] Pandita R, Chakraborty A, Seiler P, and Balas G, "Reachability and Region of Attraction Analysis Applied to GTM Dynamic Flight Envelope Assessment," in *AIAA Guidance, Navigation, and Control Conference*, 2009 [Online]. Available: http://dx.doi.org/10.2514/6.2009-6258

[10] Zheng, W., Li, Y., Zhang, D., Zhou, C., and Wu, P., "Envelope protection for aircraft encountering upset condition based on dynamic envelope enlargement," *Chinese Journal of Aeronautics*, vol. 31, no. 7, pp. 1461–1469, Jul. 2018. DOI: 10.1016/j.cja.2018.05.006, [Online].

[11] Yuan, G., and Li, Y., "Determination of the flight dynamic envelope via stable manifold," *Measurement and Control*, vol. 52, no. 3–4, pp. 244–251, Feb. 2019. DOI: 10.1177/0020294019830115, [Online].

[12] Lombaerts, T., Schuet, S., Wheeler, K., Acosta, D., and Kaneshige, J., "Robust Maneuvering Envelope Estimation Based on Reachability Analysis in an Optimal Control Formulation," *Conference on Control and Fault-Tolerant Systems* (SysTol), IEEE Publ., Piscataway, NJ, Oct. 2013, pp. 318–323. DOI: 10.1109/SysTol.2013.6693856, [Online].

[13] Harno, H. G., Kim, Y., "Safe Flight Envelope Estimation for Rotorcraft: A Reachability Approach," *18th International Conference on Control, Automation, Systems* (ICCAS), IEEE Publ., Daegwallyeong, South Korea, Dec. 2018.

[14] Tang, L., Roemer, M., Ge, J., Crassidis, A., Prasad, J., and Belcastro, C., "Methodologies for Adaptive Flight Envelope Estimation and Protection," *AIAA Guidance, Navigation, and Control Conference*, AIAA Paper 2009-6260, Aug. 2009.

[15] Oort, E. V., Chu, P., and Mulder, J. A., "Maneuvering Envelope Determination Through Reachability Analysis," *Advances in Aerospace Guidance, Navigation and Control*, edited by Holzapfel, F., and Theil, S., Springer–Verlag, Heidelberg, 2011, pp. 91–102. DOI: 10.1007/978-3-642-19817-5_8, [Online].

[16] De Marco A, Duke E, and Berndt J, "A General Solution to the Aircraft Trim Problem," in *AIAA Modeling and Simulation Technologies Conference and Exhibit*, 2007 [Online]. Available: http://dx.doi.org/10.2514/6.2007-6703

[17] Kampen E, Chu QP, Mulder JA, and Emden MH, "Nonlinear aircraft trim using internal analysis," in *AIAA Guidance, Navigation and Control Conference*, 20-23 Aug. 2007.

[18] Goman MG, Khramtsovsky AV, and Kolesnikov EN, "Evaluation of Aircraft Performance and Maneuverability by Computation of Attainable Equilibrium Sets," *Journal of Guidance, Control, and Dynamics*, vol. 31, no. 2, pp. 329–339, Mar. 2008. DOI: 10.2514/1.29336, [Online].

[19] Yinan L, Lingyu Y, and Gongzhang S, "Steady maneuver envelope evaluation for aircraft with control surface failures," in *2012 IEEE Aerospace Conference*, 2012 [Online]. Available: http://dx.doi.org/10.1109/AERO.2012.6187317

[20] Strube MJ, Sanner R, and Atkins E, "Dynamic Flight Guidance Recalibration After Actuator Failure," in *AIAA 1st Intelligent Systems Technical Conference*, 2004 [Online]. Available: http://dx.doi.org/10.2514/6.2004-6255

[21] Strube MJ, "Post-failure trajectory planning from feasible trim state sequences," M.S. thesis, Dept. Aerospace Eng. Univ. Maryland, College Park, MD, 2005.

[22] Choi HJ, Atkins E, and Yi G, "Flight Envelope Discovery for Damage Resilience with Application to an F-16," in *AIAA Infotech@Aerospace 2010*, 2010 [Online]. Available: http://dx.doi.org/10.2514/6.2010-3353

[23] Yi G, Zhong J, Atkins E, and Wang C, "Trim State Discovery with Physical Constraints," Journal of Aircraft, vol. 52, no. 1, pp. 90–106, Jan. 2015. DOI: 10.2514/1.C032619, [Online].

[24] Tang Y, Atkins E, and Sanner R, "Emergency Flight Planning for a Generalized Transport Aircraft with Left Wing Damage," in *AIAA Guidance, Navigation and Control Conference and Exhibit*, 2007 [Online]. Available: http://dx.doi.org/10.2514/6.2007-6873

[25] Asadi D, Sabzehparvar M, and Talebi HA, "Damaged airplane flight envelope and stability evaluation," *Aircraft Engineering and Aerospace Technology*, vol. 85, no. 3, pp. 186–198, May 2013. DOI: 10.1108/00022661311313623, [Online].

[26] Asadi D, Sabzehparvar M, Atkins E, and Talebi HA, "Damaged Airplane Trajectory Planning Based on Flight Envelope and Motion Primitives," *Journal of Aircraft*, vol. 51, no. 6, pp. 1740–1757, Nov. 2014. DOI: 10.2514/1.C032422, [Online].

[27] Nabi, H. N., Lombaerts, T., Zhang, Y., van Kampen, E., Chu, Q. P., de Visser, C. C., "Effects of Structural Failure on the Safe Flight Envelope of Aircraft," *Journal of Guidance, Control, and Dynamics*, vol. 41, no. 6, Jun. 2018, pp. 1257–1275. DOI: 10.2514/1.G003184, [Online].

[28] Zhang, Y., de Visser, C. C., and Chu, Q. P., "Online Safe Flight Envelope Prediction for Damaged Aircraft: A Database-Driven Approach," *AIAA Modeling and Simulation Technologies Conference*, AIAA SciTech, AIAA Paper 2016-1189, 2016. DOI: 10.2514/6.2016-1189, [Online].

[29] Zhang, Y., de Visser, C. C., and Chu, Q. P., "Aircraft Damage Identification and Classification for Database-Driven Online Safe Flight Envelope Prediction," *AIAA Atmospheric Flight Mechanics Conference*, AIAA SciTech, AIAA Paper 2017-1863, 2017. DOI: 10.2514/6.2017-1863, [Online].

[30] Zhang, Y., de Visser, C. C., Chu, Q. P., "Database Building and Interpolation for a Safe Flight Envelope Prediction System," *2018 AIAA Information Systems-AIAA Infotech @ Aerospace*, Jan. 2018. DOI: 10.2514/6.2018-1635, [Online].

[31] Foster J, Cunningham K, Fremaux C, Shah G, Stewart E, Rivers R, et al., "Dynamics Modeling and Simulation of Large Transport Airplanes in Upset Conditions," in *AIAA Guidance, Navigation, and Control*

*Conference and Exhibit*, 2005 [Online]. Available: http://dx.doi.org/10.2514/6.2005-5933
[32] "GTM_DesignSim," v1308, [Online]. Available: https://github.com/nasa/GTM_DesignSim
[33] Gregory I, Cao C, Xargay E, Hovakimyan N, and Zou X, "L1 Adaptive Control Design for NASA AirSTAR Flight Test Vehicle," in *AIAA Guidance, Navigation, and Control Conference*, 2009 [Online]. Available: http://dx.doi.org/10.2514/6.2009-5738
[34] Jordan T, Foster J, Bailey R, and Belcastro C, "AirSTAR: A UAV Platform for Flight Dynamics and Control System Testing," in *25th AIAA Aerodynamic Measurement Technology and Ground Testing Conference*, 2006 [Online]. Available: http://dx.doi.org/10.2514/6.2006-3307
[35] Jordan T, Langford W, and Hill J, "Airborne Subscale Transport Aircraft Research Testbed - Aircraft Model Development," in *AIAA Guidance, Navigation, and Control Conference and Exhibit*, 2005 [Online]. Available: http://dx.doi.org/10.2514/6.2005-6432
[36] Murch AM, "Aerodynamic modeling of post – stall and spin dynamics of large transport airplanes," M.S. thesis, School Aerospace Eng., Georgia Inst. Tech., Atlanta, GA, USA, 2007.
[37] Shah G., "Aerodynamic Effects and Modeling of Damage to Transport Aircraft," in *AIAA Atmospheric Flight Mechanics Conference and Exhibit*, 2008 [Online]. Available: http://dx.doi.org/10.2514/6.2008-6203
[38] Stevens B, Lewis F, and Johnson E, *Aircraft control and simulation*, 3rd ed. Hoboken: Wiley, 2016.
[39] Khalil H, *Nonlinear systems*, 3rd ed., Upper Saddle River, N.J., USA: Prentice Hall, 2002.
[40] Lombaerts T, Schuet S, Acosta D, and Kaneshige J, "On-Line Safe Flight Envelope Determination for Impaired Aircraft," in *Advances in Aerospace Guidance, Navigation and Control*, Springer International Publishing, pp. 263–282, 2015. DOI: 10.1007/978-3-319-17518-8_16, [Online].
[41] Ouellette JA, "Flight dynamics and maneuver loads on a commercial aircraft with discrete source damage," M.S. thesis, Aerospace Eng., Virginia Polytech. Inst. State Univ., Blacksburg, VA, USA, 2010.
[42] [dataset] Norouzi R, Kosari A, Sabour MH, "Data for: Maneuvering Flight Envelope Evaluation and Analysis of Generic Transport Model with Control Surfaces Failures", Mendeley Data, v1, 2018. http://dx.doi.org/10.17632/k4ntmx43x5.1
[43] Edwards C, Lombaerts T, and Smaili H, *Fault tolerant flight control*. Berlin: Springer, 2010, ch. 14, sec. 3.6.
[44] Philips W, *Mechanics of Flight*, Hoboken: Wiley, 2004, p. 564.
[45] McClamroch NH, *Steady Aircraft Flight and Performance*, Princeton, NJ, USA: Princeton University press, 2011.
[46] Ouellette JA, Raghavan B, Patil M, and Kapania R, "Flight Dynamics and Structural Load Distribution for a Damaged Aircraft," in *AIAA Atmospheric Flight Mechanics Conference*, 2009 [Online]. Available: http://dx.doi.org/10.2514/6.2009-6153